\documentclass[11pt]{article}
\usepackage[utf8]{inputenc}
\usepackage{xspace}
\usepackage{epsfig,graphicx}
\usepackage{amsmath,amssymb}
\usepackage{theorem}
\usepackage{setspace}
\usepackage{multirow}
\usepackage{fancyhdr} 
\usepackage[numbers,sort]{natbib} 
\usepackage{empheq}
\usepackage{cleveref}

\usepackage{tikz}
\usepackage{pgfplots}
\usepackage{cancel}
\usepackage{enumerate}

\onehalfspace

\usepackage[algo2e,lined,algoruled,linesnumbered,titlenumbered]{algorithm2e}
\allowdisplaybreaks[1]

\SetKwRepeat{Do}{do}{while}%

\setcitestyle{authoryear,open={(},close={)}}

\fancypagestyle{memo}{
	\pagestyle{fancy}
	\fancyhf{}%
	\fancyhead[RO]{\thepage}%
	\fancyhead[LO]{\rightmark}%
	
}%

\theoremstyle{plain}
\newtheorem{theorem}{Theorem}

\newtheorem{coro}{Corollary}

\newtheorem{example}{Example}

\theoremstyle{plain} {
	\theorembodyfont{\rmfamily}%

}

\newcommand{\ProofNoNL}{{\bf \noindent Proof.}\xspace}

\newcommand{\EndProofNoNL}{\hfill $\Box$ \par \bigskip}

\newcommand{\wloge}{without loss of generality\xspace}
\newcommand{\be}{{\ensuremath{\bar e}}\xspace}

\newcommand{\Real}{\ensuremath{\mathbb{R}}\xspace}
\newcommand{\N}{\ensuremath{\mathbb{N}}\xspace}

\makeatletter
\renewcommand*\env@matrix[1][c]{\hskip -\arraycolsep
	\let\@ifnextchar\new@ifnextchar
	\array{*\c@MaxMatrixCols #1}}
\makeatother

\usepackage[normalem]{ulem}

\definecolor{verde}{rgb}{0,0.4,0}
\definecolor{verdeclaro}{rgb}{0.01, 0.75, 0.24}
\begin{document}
\title{On the complexity of the upgrading version of the Maximal Covering Location Problem}
\author{\smaller Marta Baldomero-Naranjo$^1$\footnote{ Corresponding author: martbald@ucm.es (M. Baldomero-Naranjo)}, J\"org Kalcsics$^2$, Antonio M. Rodr\'iguez-Ch\'ia$^3$ \\[1ex]
\smaller $^1$ Departamento de Estad\'istica y Ciencia de los Datos, Universidad Complutense de Madrid, \\ \smaller Facultad de Estudios Estadísticos, Madrid, 28040, Spain,  martbald@ucm.es, 0000-0002-0788-7412 \\
\smaller $^2$ School of Mathematics, University of Edinburgh, Edinburgh, EH9 3FD, United Kingdom, \\ \smaller joerg.kalcsics@ed.ac.uk, 0000-0002-5013-3448\\
\smaller $^3$ Departamento de Estad\'istica e Investigaci\'on Operativa, Universidad de C\'adiz, Facultad de Ciencias, \\ \smaller Puerto Real (C\'adiz), 11510, Spain,  antonio.rodriguezchia@uca.es, 0000-0002-1708-2108\\
}
\date{\smaller \today}

\maketitle

\begin{abstract}

	In this paper, we study the complexity of the upgrading version of the maximal covering location problem with edge length modifications on networks. This problem is NP-hard on general networks. However, in some particular cases, we prove that this problem is solvable in polynomial time. The cases of star and path networks combined with different assumptions for the model parameters are analysed.
	
	In particular, we obtain that the  problem on star networks is solvable in $O(n\log n)$ time for uniform weights and  NP-hard for non-uniform weights.  On paths, the single facility problem is solvable in $O(n^3)$ time, while the $p$-facility problem is NP-hard even with uniform costs and upper bounds (maximal upgrading per edge), as well as, integer parameter values. 
	Furthermore, a pseudo-polynomial algorithm is developed for the single facility problem on trees with integer parameters.

\end{abstract}

\section*{Statements and Declarations}
The authors do not have any financial and personal relationships with other people or organizations that could inappropriately influence (bias) our work.
\section{Introduction}
The Maximal Covering Location Problem (MCLP) is a well-known location problem ana\-lysed by several researchers in the last decades since its introduction in \cite{ChuRev74}. The aim of this problem is, given a set of demand points, to locate a fixed number of facilities to maximise the amount of covered demand.  A demand point is considered to be covered if its distance to a facility is smaller than or equal to a given coverage radius.

Since the seminal paper by \cite{ChuRev74}, many variants of this problem have been analysed depending on the supporting space, e.g., recently in the discrete version \citep{MarMarRodSal18,GarMar19,AraBlaFer20,CorFurLju19}, on networks \citep{BalKalRod20,FroMaiHam20,BerKalKrass16}, and in a continuous space \citep{BlaGaz21,YanXiaZhanZhouYanXu20, BanKia17}.
 
In this paper, we follow an algorithmic approach to deal with the \emph{upgrading} version of the MCLP (Up-MCLP). The upgrading maximal covering location problem aims to find the best locations for $p$ service facilities and the edge length reduction within the given budget to cover the maximum demand. In this context, upgrading an edge means reducing its length (within certain limits) at a given cost which is proportional to the extent of the reduction. Thus, it has to be decided on which edges to invest (allocate part of the budget) in order to maximise the coverage. This problem has several practical applications.

One real-life application of this problem is when a government wants to enhance the accessibility of public services for its citizens, such as healthcare centers, educational institutions, or social welfare facilities. Since improving accessibility is closely related to distances, one approach to achieving this goal is to invest in infrastructure to reduce travel times to these services. This typically involves a combination of opening new facilities and improving transportation infrastructure, e.g., upgrading roads to highways, adding new lanes, and enhancing public transportation by incorporating high-speed lines, dedicated bus routes, and increasing service frequency.

In the private sector, telecommunication, gas, and electricity  companies face a similar challenge when they aim to expand their coverage. Further application of this problem can be found in areas like shopping centers or airports. The objective is to strategically position services such as defibrillators, ATM, information posts, etc., while simultaneously building additional passenger conveyors or escalators. The goal is to ensure that the maximum number of people are within a reasonable walking distance of these facilities.

As a consequence of its wide range of applications, studying the upgrading version of classical problems is a recent trend that interests many researchers. The motivation of upgrading problems is that in some applications, the parameters of a network can be modified so as to obtain better solutions. Two key parameters of a network are demand weights and edge (arc) lengths, and allowing to adjust them means that they are decision variables of the model. Accordingly, we can consider different versions of upgrading, as follows. 
\begin{itemize}
    \item Upgrading nodes considering that the weight of the nodes can be modified (usually decreased) subject to a prespecified budget. One possible application is the problem of locating a hospital to cover the demand of the cities of a region. The demand in each city for hospital services is represented by a weight and the demand may be reduced if a primary care center is located in the city (which involves a cost and there is a budget for this purpose). The following are a few examples of problems where this version has been applied: the 1-median problem \citep{Gassner07UpDown1median}, the 1-center problem \citep{Gassner09UpDown1center}, the Euclidean 1-median problem \citep{Plas16}, the $p$-median problem \citep{sepasian15uppmedian}, and the network delay minimization problem \citep{MedBogSin18}. 
\item Upgrading nodes such that the length of all edges incident to this node is reduced. One of the applications is to improve the access to a facility subject to a cost. This type of upgrading has recently been studied in relation to the $p$-center problem \citep{AntLanSad23}. In this paper, the authors do not consider an underlying network and its individual edges, but instead, they only consider direct connections between customers and facilities as a whole and upgrade those (additionally, upgrading a connection for one customer has no effect on the connection of another customer). This type of upgrading has also been applied in other combinatorial optimization problems, such as: the spanning tree problem \citep{ALVAREZMIRANDA201713}, in communication and signal flow problems \citep{PaiSah95}, and in the hub-location problem \citep{BlaMar19}, among others.
    \item Discrete upgrades over edges (arcs) considering that the edge (arc) length can be reduced incrementally in fixed steps, with each step reducing the length by a (potentially) different amount and at a different cost. For example, each step could reduce the edge length by 10\%, up to a maximum of 50\%, with each upgrade step becoming progressively more expensive to implement. If there is just one step, then the decision is merely whether or not to upgrade the edge (arc). This type of upgrading has recently been studied in relation to the $p$-center problem with direct connection upgrades \citep{AntLanSad23}.  Discrete upgrades have also been applied for other combinatorial optimization problems, such as: the minimum cost flow problem \citep{BuKoKirTho17}, the minimization of the maximum travel time \citep{CamLoZha06}, the accessibility arc problem \citep{MaCoGoSoSpi13}, and the Graphical TSP \citep{LanPla23}, among others. 
\item Continuous upgrades over edges (arcs) considering that the length of the edges (arcs) can be modified by any increment subject to a prespecified budget.
For covering problems, this has recently been considered by \cite{BalKalMarRod22}, who proposed several mixed-integer programming formulations to solve the problem exactly. Other facility location problems where this approach has been studied include:  the 1-center problem \citep{Sepasian181center}, the $p$-center problem with direct connection upgrades \citep{AntLanSad23}, and the $p$-median problem \citep{AfrAliBar20,EspMar23}. Also, this has been applied in combinatorial optimization problems, such as: the min-max spanning tree problem \citep{SeMon17}, the minimum flow cost problems \citep{DemNolWirUpMinFlow}, the maximal shortest path interdiction problem \citep{ZhaGuaPar21}.  
\end{itemize}

To the best of our knowledge, this is the first attempt to analyse the complexity of the MCLP with continuous edge upgrades for different types of graphs. 
It is known that the MCLP is NP-hard on a general network even without upgrades \citep{MeZeHa83}. Therefore, the Up-MCLP is also NP-hard on a general network, since the MCLP is a special case of the Up-MCLP that has the maximal upgrade per edge equal to zero for all edges. However, the classical MCLP is polynomial-time solvable on trees \citep{MeZeHa83} and consequently also on stars and paths \citep{hasTam91}. The open question we address in this work is whether it is possible to prove similar results for the Up-MCLP. 

Therefore, in this paper, we turn our attention to particular networks such as stars, paths, and trees and study the complexity of the Up-MCLP and provide polynomial and pseudo-polynomial time algorithms under different assumptions for the parameters. In particular, we prove the following results. 
\begin{itemize}
    \item On star networks, the $p$-facility problem is polynomial time solvable for uniform weights, while for non-uniform weights already the single facility problem is NP-hard even with uniform costs and upper bounds as well as integer parameter values. 
    \item On paths, the single facility problem is solvable in polynomial time, while the $p$-facility problem is NP-hard even with uniform costs and upper bounds as well as integer parameter values.
    \item On trees, the single facility weighted problem with integer parameters has a pseudo-polynomial algorithm; note that this problem is NP-hard (as a
star is also a tree).
 \end{itemize}    
   Similar results have been obtained for various location problems, 
   including covering problems \citep{MeZeHa83,BhaNan13,hartmann2021continuous,hasTam91}, center problems \citep{wang2021n}, median problems \citep{OudStal21,AfrAliBar19median}, or other facility location problems \citep{nguTeh21,PuRiSco18,LaPuRiCco18,KalNiPuRod15}, as well as flow/shortest path problems  \citep{SuAlAr21,DanVasCve21,Hopp22}. 
    

A summary of the obtained results can be found in Table~\ref{tab:summary}, where the first column indicates the complexity of the Up-MCLP, the second column shows the graph type, the third column points out the number of facilities, the three following columns provide information on the parameters of the problem: uniformity of weights (unif.\ $w$), upgrading costs (unif.\ $c$), and upper bounds on the amount of upgrade on an edge (unif.\ $u$), respectively. The last column indicates the theorem in which the result is proven. Note that ``Y" 
represents that the uniformity on the respective parameters is assumed, ``N" 
represents that the non-uniformity on the respective parameters is assumed, and blank space means that no assumptions are made on the corresponding uniformity of the parameters.

	\begin{table}[htbp]
 		\centering
 	
 		\begin{tabular}{ccccccc}
 			\hline
 			\hline
 			Complexity & Graph & Num. facilities & Unif.\ $w$ & Unif.\ $c$ & Unif.\ $u$ & Result\\
 			\hline
 			\hline
 			\multirow{2}{*}{Polynomial } & Star  & $p$     & Y    &       & & Theorem~\ref{theo:NPhard:1_facility_stars1_uni} \\
 			\cline{2-7}          & Path  & 1     &       &       &  & Theorem~\ref{theo:complexity:1_facility_path}\\
 			\hline
 			\hline
 			\multirow{2}{*}{NP-hard} & Star  & 1     &N    &  Y       & Y   & Theorem~\ref{theo:NPhard:1_facility_stars1_nonuni}\\
 			\cline{2-7}          & Path  & $p$     &   N     & Y     & Y & Theorem~\ref{theo:NPhard:p_facility_path} \\
 			\hline
 			\hline
 		\end{tabular}%
 		\caption{Summary of complexity results.}
 		\label{tab:summary}%
 	\end{table}%

The rest of the paper is structured as follows. In Section~\ref{sec:ProblemDescription} the problem is introduced. Section~\ref{sec:star} analyses the  problem on star networks, a polynomial time algorithm is derived for uniform weights and its NP-hardness is proven for non-uniform weights. In Section~\ref{sec:facility:path}, a polynomial time algorithm is presented for the single facility problem on path networks. Moreover, we prove that the $p$-facility problem is NP-hard even with uniform costs and upper bounds and integer parameters values. In Section~\ref{sec:tree} 
a pseudo-polynomial algorithm is presented for the single facility problem on tree networks with integer parameters. Lastly, our conclusions are presented in Section~\ref{sec:conclusion}.

\section{Definitions and Problem Description}
\label{sec:ProblemDescription}%
Let $N=(V,E,\ell)$ be an undirected network with node set $V=\{1, \ldots, n\}$ and edge set $E$, where $|E|=m$. Slightly abusing notation, sometimes we denote the nodes also as $v_1,\ldots,v_n$ instead of $1,\ldots,n$.
Every edge $e=(k,q)=(q,k)\in E$, $k,q \in V,$ has a positive length $\ell_e=\ell_{(k,q)}$. For $i,j\in V,$ $d(i,j)$ is the length of the shortest path connecting $i$ with $j$. Furthermore, we are given a fixed coverage radius $R > 0$. We say that a node $i \in V$ is \emph{covered} by a facility at node $j$ if $d(i,j) \leq R.$ Finally, for each node $i \in V$ we are given a non-negative \emph{weight} $w_i$ that specifies the demand at the node. 

The length $\ell_e$ of each edge $e\in E$ can be reduced by an amount lower than or equal to $u_e$ $\in [0,\ell_e)$, $e\in E$. Without loss of generality, we assume that $\ell_e-u_e\leq R$, for $e\in E$ (if that were not the case, i.e., there was an edge $e \in E$ such that $\ell_{e}-u_{e}> R$, then $e$ can be removed from the network without affecting the optimal solution). 
Moreover, reducing the length of edge $e \in E$ by an amount $\delta_e \in [0, u_e]$ comes at a positive cost of $\delta_e c_e$, for $c_e \in \mathbb{R}^+$, and there is a budget constraint $B \in \mathbb{R}^+$ on the overall cost of reduction. Again, without loss of generality, we assume that $c_eu_e\leq B$, for $e\in E$ (if that were not the case, i.e., there was a cost $c_e$ for $e\in E$ such that $c_{e}u_{e}> B$, then $u_e$ can be substituted by $u_{e}= B/c_{e}$ without affecting the optimal solution since it is not possible to pay for a reduction whose cost is greater than the budget). For short, the labels $(\ell_e,u_e,c_e)$ have been given to the edges representing the length, the maximal reduction, and the cost per unit of reduction of each edge. Finally, we assume that facilities can only be located at nodes. 

The upgrading maximal covering location problem (Up-MCLP) aims to locate $p$ service facilities covering the maximum demand taking into account that the total cost for the edge length reductions is within the given budget.

Let $\delta = (\delta_e)_{e \in E}$ denote a vector of edge length reductions, $0 \le \delta_e \le u_e$, for $e\in E$. For $i,j\in V$ and reductions $\delta$, $d(i,j,\delta)$ is the length of a shortest path between $i$ and $j$ in the network after the reductions $\delta$ have been applied. Moreover, we call $i$ reachable from $j$ (and vice versa), if we can find an upgrade $\delta$ of the edges such that $i$ is covered by $j$, i.e., $d(i,j,\delta) \le R$. Finally, for $p \in \mathbb{N}$, let $X_p \subseteq V$ denote a set of $p$ nodes, let $(X_p,\delta)$ represent a feasible solution of the Up-MCLP, and let $C(X_p,\delta) = \{i \in V \mid \exists j \in X_p: d(i,j,\delta) \le R \}$ denote the set of all nodes covered by a facility in $X_p$ after the edge upgrades $\delta$.
Then, the $p$-Up-MCLP can be formulated as:
\[ \max \left\{ \sum_{i \in C(X_p,\delta)}\, w_i\: \Big|\: \sum_{e \in E} c_e \delta_e \le B, X_p \subseteq V, |X_p|=p, 0 \le \delta_e \le u_e, e \in E \right\}.
\]

For a given weight threshold $T \in \mathbb{R}^+$, we define the decision version $p$-Up-MCLP-D of $p$-Up-MCLP as:
\begin{description}
\item[Input:] Network $N=(V,E,\ell)$, number of facilities $p$, coverage radius $R$, upgrading bounds and costs $(u_e)_{e \in E}$ and $(c_e)_{e \in E}$, respectively, and budget $B$.
\item[Question:] Does there exist a set $X_p \subseteq V$ and edge upgrades $\delta=(\delta_e)_{e \in E} \in \bigtimes_{e \in E} [0,u_e]$ with $|X_p|=p$ and $\sum_{e \in E} c_e \delta_e \le B$ such that $\sum_{i \in C(X_p,\delta)}\, w_i \ge T$.
\end{description}

As previously discussed, the Up-MCLP is NP-hard on a general network. However, the classical MCLP is polynomial time solvable on trees and paths. Indeed, \cite{MeZeHa83} present an $O(n^2p)$ algorithm for this problem on trees and \cite{hasTam91} derive an $O(np)$ algorithm on paths. The objective of this paper is to analyse the upgrading version of this problem in star, path, and tree networks to determine for which cases and under which assumptions on the demand weights $w_i$, upgrading costs $c_e$, or upgrading bounds $u_e$ the problem can be solved in polynomial time. 
Hereinafter, the following expressions mean: 
\begin{itemize}
  \item \emph{Uniform weights:} $w_i = w$ for all $i \in V$ and some $w \in \mathbb{R}^+$. 
  \item \emph{Uniform costs:} $c_e=c$  for all $e \in E$ and some $c \in \mathbb{R}^+$. 
  \item \emph{Uniform bounds:} $u_e=u$ for all $e \in E$ and some $u \in \mathbb{R}^+$.
\end{itemize}
In the next section, we focus on the star network case.

\section{Star networks}\label{sec:star}
 
In this section, we establish several complexity results for the problem on star networks depending on the characteristics of the input parameters. First, we present a result for the unweighted single facility problem.

\begin{theorem}
  \label{theo:NPhard:1_facility_stars1_uni}
  The single facility maximal covering problem with edge length variations, 1-Up-MCLP, can be solved in $O(n\log n)$ time for uniform weights (unweighted case) on star networks.
\end{theorem}
 
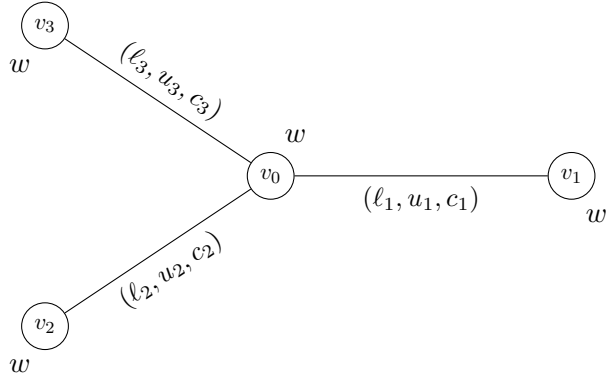
\begin{figure}[htb]
\centering
\begin{tikzpicture}
	\node[scale=.8,auto=left,circle,draw,label=  80:$w$]   (v0) at ( 0, 0) {$v_0$};
	\node[scale=.8,auto=left,circle,draw,label= -80:$w$] (v1) at ( 4, 0) {$v_1$};
	\node[scale=.8,auto=left,circle,draw,label=-100:$w$] (v2) at (-3,-2) {$v_2$};
	\node[scale=.8,auto=left,circle,draw,label=-100:$w$] (v3) at (-3, 2) {$v_3$};
			
	\draw (v0) -- (v1) node[pos=0.5, below, sloped]{\small $(\ell_1,u_1,c_1)$};
	\draw (v0) -- (v2) node[pos=0.5, below, sloped]{\small $(\ell_2,u_2,c_2)$};
	\draw (v0) -- (v3) node[pos=0.5, above, sloped]{\small $(\ell_3,u_3,c_3)$};
\end{tikzpicture}
\caption{Illustration for the uniform case in Theorem~\ref{theo:NPhard:1_facility_stars1_uni}, with edge labels $(\ell_j,u_j,c_j)$.}
\label{fig:complexity:star:uni}
\end{figure}

\ProofNoNL
Let $N=(V,E,\ell)$ be a star network with central node $v_0$ and satellite nodes $v_1,\ldots,v_n$. See Figure~\ref{fig:complexity:star:uni} for an illustration.
 
First, we can assume without loss of generality that the facility is located in the central node, $v_0$. Indeed, since the weights are uniform, locating a facility in a central node is at least as good as locating it in a satellite node as a facility located in a satellite node will
never be able to cover more nodes than one that is located in the central node.  
 
Then, we can find an optimal upgrading of the edges in polynomial time as follows. Let $\Delta_i = \ell_{(0, i)}-R$. If $\Delta_i \le 0$, then $v_i$ is automatically covered and can be discarded. If $\Delta_i > u_i$, then $v_i$ is unreachable and can also be discarded. All remaining satellite nodes are sorted by non-decreasing minimal upgrading costs, i.e., by $c_i \Delta_i$. Then, starting with the first node $v_i$ in the list, we upgrade the edge by $\delta_i = \min \{B/c_i,\, \Delta_i \}$ and update the budget $B = B - c_i\delta_i$. If $B>0$, we continue with the next node in the list. Otherwise, we stop. 
 
Indeed, this procedure provides an optimal solution. If this was not the case, then an optimal upgrade would provide a set of covered nodes $\hat V$, such that there exist $v_i \in \hat V$ and $v_k \in V \setminus \hat V$ satisfying $c_i \Delta_i > c_k \Delta_k$. However, upgrading edge $e_k$ instead of $e_i$ covers $v_k$ without decreasing the total number of covered nodes and without increasing the amount of used budget. 
Concerning the last node $v_j$ for which we upgrade the incident edge, if $B/c_j < \Delta_j$, then we cannot reduce the length of $e_j$ enough to cover $v_j$. However, as we sorted the nodes by non-decreasing minimal upgrading costs, spending the remaining budget instead on some other edge $e_k$ incident to another uncovered node $v_k$, $k \not\in C(\{v_0\}, \delta)$, will also not allow us to cover that node. Hence, spending the remaining budget on $e_j$ is as good as spending it on some other edge or even as good as not spending it at all, which concludes the argument.

The complexity of this procedure is dominated by the step of sorting the minimal upgrading costs, i.e., is $O(n\log n)$. All other steps are computed in constant time. Therefore, the unweighted 1-Up-MCLP can be solved in $O(n\log n)$ time on star networks.
\EndProofNoNL
  
 
  
After showing that the unweighted single facility problem can be solved in polynomial time, the next natural step is to study the $p$-facility case, i.e., $p$-Up-MCLP.    
     
\begin{theorem}
 	\label{theo:NPhard:p_facility_star}
 	The unweighted $p$-Up-MCLP can be solved in $O(n\log n)$ time on star networks.
\end{theorem}
\ProofNoNL
Let $N=(V,E,\ell)$ be a star network with central node $v_0$ and satellite nodes $v_1,\ldots,v_n$.
Again, we can assume \wloge that in an optimal solution one of the facilities is located in the central node $v_0$, as a facility located in a satellite node will never be able to cover more nodes than one that is located in the central node. 



We start by using the same algorithm as in the proof of Theorem~\ref{theo:NPhard:1_facility_stars1_uni} to compute optimal edge upgrades $\delta^*$ for the case of just one facility being located at $v_0$. Let $V^- = V \setminus C(\{v_0\}, \delta^*)$ be the set of nodes that are still uncovered after this upgrade. 
If $|V^-| > p-1$, then since each node has the same weight and no facility in a satellite node can cover a node that could not already be covered by $v_0$, putting facilities on $p-1$ arbitrary nodes in $V^-$ is optimal. If instead $|V^-| \le p-1$, then all nodes can be covered and the solution is trivially optimal.

Finally, for an arbitrary upgrade $\delta$ by the proof of Theorem~\ref{theo:NPhard:1_facility_stars1_uni} we have $|C(\{v_0\}, \delta)| \le |C(\{v_0\}, \delta^*)|$. Therefore, even if we distribute the remaining $p-1$ facilities optimally, we will not be able to cover more nodes using $\delta$ than using $\delta^*$ (again, because facilities in satellite nodes cannot cover more nodes than $v_0$ for a given upgrade).


The complexity of this construction is dominated by the step of sorting the minimal upgrading costs, i.e., $O(n\log n)$. Therefore, the unweighted $p$-Up-MCLP can be solved in $O(n\log n)$ time on star networks.

\EndProofNoNL


\noindent  
Next, we discuss the case in which the node weights are not uniform. 
\begin{theorem}
\label{theo:NPhard:1_facility_stars1_nonuni}
The 1-Up-MCLP is NP-hard on star networks for non-uniform weights (weighted case) even assuming uniformity in upgrading costs and upper bounds as well as integrality of the input parameter values.
\end{theorem}

\ProofNoNL
We prove the result by reducing KNAPSACK to the decision problem 1-Up-MCLP-D. 
Let an instance of KNAPSACK be given with $n$ items of positive weight $g_i \in \mathbb{N}$ and positive value $b_i \in \mathbb{N}$, $1 \le i \le n$, knapsack capacity $K \in \mathbb{N}$, and target value $U > 0$. A solution (i.e., a Yes-Input) to KNAPSACK is a set $M \subseteq \{1,\ldots,n\}$ such that $\sum_{i \in M} g_i \le K$ and $\sum_{i \in M} b_i \ge U$. Without loss of generality, we assume $g_i \le K$, $1 \le i \le n$.

First, we observe that 1-Up-MCLP-D is in NP since a given solution for 1-Up-MCLP-D can be verified as such in polynomial time.
Next, given an instance for KNAPSACK, we construct an instance of the edge upgrading problem in polynomial time as follows. Let $N=(V,E,\ell)$ be a star network with central node $v_0$, satellite nodes $v_1,\ldots,v_n$, and edges $e_i = (v_0,v_i)$. Each satellite node $v_i$ corresponds to one item $i$, and the weight $w_i$ of the node equals the item value $b_i$. The central node is given weight $w_0=W > \sum_{i=1}^n b_i$. Finally, the weight threshold $T$ for 1-Up-MCLP-D is defined as $T=W+U$.

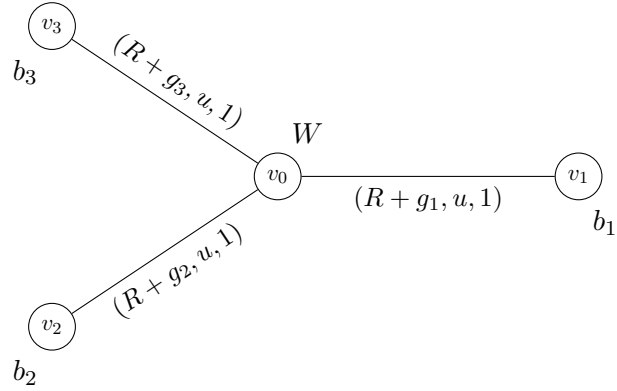
\begin{figure}[htb]
  \centering
  \begin{tikzpicture}
 	\node[scale=.8,auto=left,circle,draw,label=  80:$W$]   (v0) at ( 0, 0) {$v_0$};
 	\node[scale=.8,auto=left,circle,draw,label= -80:$b_1$] (v1) at ( 4, 0) {$v_1$};
 	\node[scale=.8,auto=left,circle,draw,label=-100:$b_2$] (v2) at (-3,-2) {$v_2$};
 	\node[scale=.8,auto=left,circle,draw,label=-100:$b_3$] (v3) at (-3, 2) {$v_3$};
 		
 	\draw (v0) -- (v1) node[pos=0.5, below, sloped]{\small $(R+g_1,u,1)$};
 	\draw (v0) -- (v2) node[pos=0.5, below, sloped]{\small $(R+g_2,u,1)$};
 	\draw (v0) -- (v3) node[pos=0.5, above, sloped]{\small $(R+g_3,u,1)$};
  \end{tikzpicture}
  \caption{Illustration for the proof of Theorem~\ref{theo:NPhard:1_facility_stars1_nonuni}, with edge labels $(\ell_j,u,c)$.}
  \label{fig:complexity:star:nonuni}
\end{figure}

Let $c$ and $u$ be the unit upgrading costs and, respectively, the upper bounds on the upgrade in each edge.  We set $c=1$, $u = \max_{i=1,\ldots,n} g_i \le K,$ $R=u+1$, and the length of each edge $e_i$ as $\ell_{i} = R+g_i$. Finally, the budget equals the upgrading cost times the knapsack capacity, i.e., $B=K$. See Figure~\ref{fig:complexity:star:nonuni} for an illustration.
 
Let $(\{x\},\delta)$ be a solution for 1-Up-MCLP-D. As $\sum_{i \in C(\{x\},\delta)} w_i \ge T > W$, the central node must be covered. Therefore, we can assume without loss of generality that $x=v_0$. Let $M = C(\{v_0\},\delta) \setminus \{v_0\}$ be the set of all satellite nodes covered by $v_0$ after the upgrade $\delta$. A satellite node $v_i$ is hereby covered if and only if $g_i\leq \delta_i \leq u$.
As 
\[ \sum_{i \in M}  \delta_i \le \sum_{i=1}^n  \delta_i \le B = K
\]
we get $\sum_{i \in M} g_i \le K$. Moreover, as 
\[ \sum_{i \in C(\{v_0\},\delta)} w_i = W + \sum_{i \in M} w_i \ge T = W+U,
\]
we obtain $\sum_{i \in M} w_i = \sum_{i \in M} b_i \ge U$ and $M$ is a solution to KNAPSACK.
Vice versa, for the same star network any solution $M$ to KNAPSACK can easily be converted into a solution for 1-Up-MCLP-D by setting $X=\{v_0\}$ and $\delta_i = g_i$, $i \in M$. 

As a result, the weighted 1-Up-MCLP-D is NP-complete on star networks and the weighted 1-Up-MCLP is NP-hard.
\EndProofNoNL
 
Consequently, the $p$-facility problem is also NP-hard. Therefore, from the analysis carried out in this section we have seen that the key feature for the Up-MCLP to be solvable or not in polynomial time on star networks is given by the uniformity or non-uniformity of weights. 
 
\section{Path networks}\label{sec:facility:path}
In this section, we analyse the complexity of Up-MCLP on path networks. First, we show that the 1-Up-MCLP is solvable in polynomial time on path networks.
 
\begin{figure}[htb]
 	\centering
 	\resizebox{.95\linewidth}{!}{
 	\begin{tikzpicture}
 		\node[scale=.8,auto=left,style=circle,draw] (n1) at (-0.5,0) {$v_s^l$};
 		\node[scale=.8,auto=left,style=circle,draw] (n2) at (2,0)  {$v_2^l$};
 		\node[scale=.8,auto=left,style=circle,draw] (n3) at (4,0)  {$v_1^l$};
 		\node[scale=.8,auto=left,style=circle,draw] (n4) at (6.5,0)  {$v_0$};
 		\node[scale=.8,auto=left,style=circle,draw] (n5) at (9,0)  {$v_1^{r}$};
 		\node[scale=.8,auto=left,style=circle,draw] (n6) at (11,0) {$v_2^{r}$};
 		\node[scale=.8,auto=left,style=circle,draw] (n7) at (13,0) {$v_t^{r}$};
 		\draw [dashed] (n1) -- (n2); 
 		\draw (n2) -- (n3) node[pos=0.5, above]{$e_2^l$};
 		\draw (n3) -- (n4) node[pos=0.5, above]{$e_1^l$}
 	node[pos=0.5, below]{$(\ell_1^l,u_1^l,c_1^l)$};
 		\draw (n4) -- (n5) node[pos=0.5, above]{$e_1^r$};
 		\draw (n5) -- (n6) node[pos=0.5, above]{$e_2^r$};
 	\draw [dashed] (n6) -- (n7); 
 	\end{tikzpicture}
}
 	\caption{Notation used in Section~\ref{sec:facility:path}}
 	\label{fig:path:1}
\end{figure}
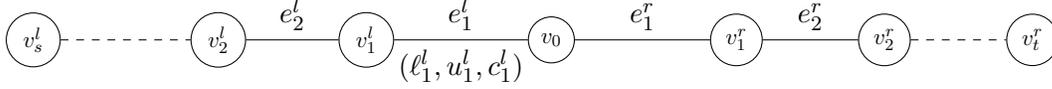

 
The resolution process consists of checking all possible facility locations, i.e., all nodes of the path, and then calculating the resulting maximal coverage that can be obtained by upgrading edges.
For each $i\in V$, we locate the facility on node $i$ of the path. We denote this node in the following as $v_0$.
Let $V^l$ and $E^l$ ($V^r$ and $E^r$) be the set of nodes and edges on the left-hand side (on the right-hand side) of $v_0$, respectively. We number the nodes and edges in $V^l$ and $E^l$ consecutively, starting with $v_1^l$ and $e_1^l$ being adjacent and, respectively, incident to $v_0$. Let, $V^l = \{v_1^l, v_2^l, \ldots, v_s^l\}$ and $E^l=\{e_1^l,e_2^l,\ldots, e_s^l\}$, where $v_s^l$ is the left most node in $V^l$. Analogously, we define $V^r = \{v_1^r, \ldots, v_t^r\}$ and $E^r = \{e_1^r, \ldots, e_t^r\}$ for the right-hand side, where $v_t^r$ is the right most node. Accordingly, we adjust the notation for the edge lengths, upgrading costs, and edge length reduction bounds.
A sketch of this notation is depicted in Figure \ref{fig:path:1}.
 
Without loss of generality, we consider only nodes in $V^l$ and $V^r$ that are reachable from $v_0$. For example for the left-hand side, $v_q^l$, $1 \le q \le s$, is reachable if there exists a vector $\delta^l = (\delta_1^l, \ldots, \delta_q^l)$, $0 \le \delta_k^l \le u_k^l$, $1\le k \le q$, such that $d(v_q^l,v_0) \le R + \sum_{k=1}^{q} \delta_k^l$ and $\sum_{k=1}^{q} c_k^l \delta_k^l \le B$.
We denote the minimal upgrading cost to make node $v_q^l$ reachable by $B_{q}^l$ which is computed by solving the following linear program whenever this value is smaller than or equal to $B$. 
 \begin{alignat}{4} 
		&\mbox{(PL-$B_{q}^l$)}&\quad& \mbox{ min }&&  \sum_{k=1}^q c_k^l\, \delta_k^l & & \nonumber \\
		 &&&\mbox{ s.t.} &   &  \sum_{k=1}^q ( \ell_k^l - \delta_k^l ) \leq R,  & \quad &   \nonumber \\
		   	&&& & & 	   0 \le \delta_k^l \le u_k^l, && k =1, \ldots, q. \nonumber 
\end{alignat}
Analogously, we define $B_q^r$ for node $v_q^r$ and these values can again be calculated by solving a linear program. Later, we will also provide a procedure to compute them efficiently by inspection.
 
With the facility located in $v_0$, we now have to decide how much of the budget $B$ we allocate to the sub-path to the left and how much to the sub-path to the right of $v_0$. We observe that in order to be able to cover a node $v_q^l$ on the left-hand side, we have to allocate at least the budget  $B_q^l$ to the left. Increasing the budget beyond $B_q^l$ will obviously only have an effect, if we increase it up to $B_{q+1}^l$. Hence, the only useful budget allocations for the left are from the set $\{B_1^l,\ldots, B_s^l\}$. We point out that the budget values are strictly increasing, the further we go to the left. Analogously we calculate these values for the right-hand side.
 
This observation allows us to develop an efficient procedure for calculating the maximal coverage we can achieve for a facility at $v_0$. We start with the left most node $v_s^l$. To cover it, we have to allocate $B^l = B_s^l \le B$ of the budget to the left-hand side. This will also cover all nodes $v_1^l,\ldots, v_{s-1}^l$, so the total coverage is $\sum_{q \in V^l} w_q^l$. The remaining budget, $B^r = B - B^l$ can be invested in the sub-path on the right of $v_0$ and having computed the values $\{B_1^r,\ldots, B_t^r\}$, it is easy to determine the right most node $v_q^r$ that can still be covered, i.e., for which $B_q^r \le B^r$ and $B_{q+1}^r > B^r$, and the resulting weight: $\sum_{k = 1}^{q} w_k^r$. This gives us the maximal coverage we can achieve under the assumption that $v_s^l$ is covered. Now we can simply repeat this procedure from 
node $v_{s-1}^l$, 
to node $v^0$, i.e., the case of $B^l=0$. Note that this already considers all possible combinations of coverage and there is no need to repeat the same steps for the nodes to the right of $v_0$.
We summarize the steps in Algorithm~\ref{algo:path}, where we denote $B_0^l=0$ ($B_0^r=0$).
 
 \begin{algorithm2e}[ht]
 	\DontPrintSemicolon
 	\SetAlFnt{\small\sl} 
 	\SetAlCapFnt{\small\sl} \AlCapFnt 
 	
 	\caption{Optimal algorithm for the single facility Up-MCLP on paths.}
 	\label{algo:path}
 	\BlankLine
 	
 	\KwIn{Network $N=(V,E,\ell)$; bounds $u_e$ and upgrade costs $c_e$, $e\in E$; coverage radius $R > 0$; budget $B>0$.}
 	\BlankLine
 	\KwOut{Optimal solution $(v^*, w^*,\delta^*)$.}
 	\BlankLine \BlankLine
 	
 	Set $w^* = 0$.
 	\BlankLine
 	
 	\ForEach{$v_0 \in V$}{
 		Determine $V^l$, $E^l$, $V^r$, $E^r$,
 		$\{B_1^l, \ldots, B_s^l\}$, and $\{B_1^r, \ldots, B_t^r\}$. \;
 		
 		\For{$q=s,\ldots,0$}{
 			Let $B^l=B_q^l$ and $B^r = B - B^l$.\;
 			Determine the maximal $1\le k \le t$ such that $B_k^r \le B^r$.\;
 			Calculate $w = \sum_{i = 1}^{q} w_i^l + \sum_{i = 1}^{k} w_i^r$. 
 			\BlankLine
 			
 			\If{$w > w^*$}{
 				Set $w^* = w,$ $\delta^*=\delta$, and $v^* = v_0$.
 			}
 		}
 	}
 	
 	\BlankLine
 	
 	\Return\ $(v^*,w^*,\delta^*)$.\;
\end{algorithm2e}
 \medskip
 
\begin{theorem} 	\label{theo:complexity:1_facility_path}
 The 1-Up-MCLP can be solved in $O(n^3)$ time on path networks. 
\end{theorem}
  
\ProofNoNL
Let $v_0$ be fixed and consider Algorithm~\ref{algo:path}. In Step~3 we have to calculate the set of reachable nodes to the left and right of $v_0$, together with the corresponding minimal upgrading cost. To do this efficiently, we start with the adjacent node $v_1^l$ to the left of $v_0$ and calculate $B_1^l$. If $B_1^l \ge B$, we can already stop. If $B_1^l= B$ we set $s=1$ and $V^l=\{v_1^l\}$, otherwise we set $s=0$ and $V^l=\varnothing$. If instead $B_1^l < B$, then we calculate $B_2^l$ for node $v_2^l$. We repeat this procedure until $B_q^l \ge B$ or we reach the left most node of the left sub-path. The sets and values to the right of $v_0$ are calculated analogously.
 
Although, $B_q^l$, $1 \le q \le s$, can be computed by solving \mbox{(PL-$B_{q}^l$)}, we next describe a procedure with lower complexity. 
First, we sort the upgrade costs $c_k^l$, $1 \le k \le q$, in non-decreasing order:
\[ c_{(1)}^l \le c_{(2)}^l \le \ldots \le c_{(q)}^l \,.
\]
Let $\Delta = d(v_q^l,v^0)-R$. If $\Delta \le 0$, then $B_q^l = 0$ and we are done. Thus, assume now that $\Delta > 0$. Pick the edge $e_{(1)}^l$ with the smallest upgrade cost $c_{(1)}^l$ and set $\delta_{(1)}^l = \min\{\Delta, u_{(1)}^l\}$, i.e., we upgrade the edge until we reach the upper bound or until we can cover $v_q^l$. Set $\Delta = \Delta - \delta_{(1)}^l$. If $\Delta = 0$, then $B_q^l = c_{(1)}^l\delta_{(1)}^l$ and we are done. Otherwise, we continue with the second-cheapest edge $e_{(2)}^l$. If we reach the last edge $e_{(q)}^l$ and still $\Delta > 0$, then $v_q^l$ cannot be covered from $v_0$, i.e., is unreachable.
The values on the right-hand side can be determined analogously.
Concerning the complexity of this step, for a given $v_q^l$, it takes $O(q \log q)$ time to sort the edges by non-decreasing upgrade costs from scratch. However, by starting with $v_1^l$ and then working our way to the left, we can calculate the sorted list incrementally, by just inserting $e_q^l$ in $O(\log q)$ time.
Going through the sorted list and checking against $\Delta$ and updating it can then be done in time proportional to $q$, i.e., in $O(q)$ time. As we can have up to $n-1$ nodes on the left-hand side, the total complexity of Step~3 is $O(\sum_{q=1}^{n-1} (\log q + q)) = O(n^2)$.

Step~5 requires constant time. The sum in Step~7 can be updated incrementally, resulting in a constant effort.
Steps~8-10 are also constant.
Finally, as $B_q^r$ is strictly increasing and, hence, we have to look for the maximal $k$ only to the right of the maximal $k$ of the previous iteration, the complexity of Step~6 for all $q=s,\ldots,0,$ is $O(n)$.
In total, the effort for Steps~4-11 is $O(n)$.
 
Thus, the total complexity of the algorithm is dominated by Step~3, i.e., $O(n^2)$. Since it needs to be repeated $n$ times, the complexity of the algorithm is $O(n^3)$.
\EndProofNoNL


At this point, we have shown that the 1-Up-MCLP on paths can be solved in polynomial time. This raises the question about the $p$-facility problem on those networks.

\begin{theorem}
\label{theo:NPhard:p_facility_path}
The $p$-Up-MCLP is NP-hard on path networks even assuming uniformity in upgrading costs and upper bounds as well as integrality in the input parameter values.
\end{theorem}
 
\ProofNoNL
We again prove the result by reducing KNAPSACK to the decision problem $p$-Up-MCLP-D. 
Let an instance of KNAPSACK be given with $n$ items of positive weight $g_i \in \mathbb{N}$ and positive value $b_i \in \mathbb{N}$, $1 \le i \le n$, knapsack capacity $K \in \mathbb{N}$, and target value $U > 0$.
We construct an instance of the upgrading problem in polynomial time as follows.

First, we observe that $p$-Up-MCLP-D is in NP since a given solution can be verified as such in polynomial time.
Next, given an instance for KNAPSACK, we construct an instance of the edge upgrading problem in polynomial time as follows. 
Let $N=(V,E,\ell)$ be a path network with $2n$ nodes $V=\{v_1,v_2,\ldots,v_{2n}\}$. Odd indexed nodes $v_{2k-1}$ have weight $w_{2k-1}=W>\sum_{i=1}^n b_i$, and even indexed nodes $v_{2k}$ have weight $w_{2k}=b_k$, $k=1,\ldots,n$. Let $V^{odd}=\{v_{2k-1} \mid k=1,\ldots,n\}$ and $V^{even}=V \setminus V^{odd}$. Moreover, let $u=\max_{i=1,\ldots,n} g_i,$ $R=1+u,$ $c=1$, $B=K$, and $T=pW+U$.

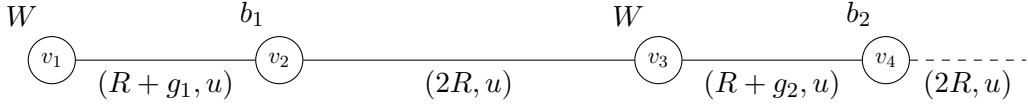
\begin{figure}[htb]
 	\centering
 		\resizebox{.95\linewidth}{!}{
		\begin{tikzpicture}
				\node[scale=.8,auto=left,circle,draw,label=100:$W$] (v1) at (0,0) {$v_1$};
				\node[scale=.8,auto=left,circle,draw,label=100:$b_1$] (v2) at (3,0)  {$v_2$};
				\node[scale=.8,auto=left,circle,draw,label=100:$W$] (v3) at (8,0)  {$v_3$};
				\node[scale=.8,auto=left,circle,draw,label=100:$b_2$] (v4) at (11,0)  {$v_4$};
				\node[scale=.8,auto=left,circle] (v5) at (13,0)  {};
				
				\draw (v1) -- (v2) node[pos=0.5, below]{$(R+g_1,u)$};
				\draw (v2) -- (v3) node[pos=0.5, below]{$(2R,u)$};
				\draw (v3) -- (v4) node[pos=0.5, below]{$(R+g_2,u)$};
				\draw [dashed] (v4) -- (v5) node[pos=0.5, below]{$(2R,u)$};
			\end{tikzpicture}
}
 	\caption{Notation used in Theorem~\ref{theo:NPhard:p_facility_path}, with edge labels $(\ell_j,u_j)$.}
 	\label{fig:path:p}
\end{figure}

Concerning the edges, for $e_{2k-1} = (v_{2k-1}, v_{2k})$ we define $\ell_{2k-1} = R + g_{k},$ for $k = 1, \ldots, n$. Moreover, for $e_{2k} = (v_{2k}, v_{2k+1})$ we define $\ell_{2k} = 2R$, for $k = 1, \ldots, n-1$.
A sketch of the resulting network is depicted in Figure~\ref{fig:path:p}.
 
Let $p = n$ and $(X_p,\delta)$ be a solution for $p$-Up-MCLP-D. As $\sum_{i \in C(X_p,\delta)} w_i \ge T > pW$, all odd-indexed nodes $v_{2k-1} \in V^{odd}$ must be covered. Therefore, we can assume without loss of generality that the $p=n$ facilities are located in the $n$ odd-index nodes, i.e., $X_p=V^{odd}$. Let $M = C(X_p,\delta) \cap V^{even}$ be the set of all even-indexed nodes covered by $X_p$ after the upgrade. An even-indexed node $v_{2k}$ is covered if and only if $g_{k} \leq \delta_{2k-1} \leq u$. As 
\[ \sum_{i \in M} \delta_i \le \sum_{i=1}^{2n} \delta_i \le B = K
\]
(recall that $c=1$) we get $\sum_{i \in M} g_i \le K$. Moreover, as 
\[ \sum_{i \in C(X_p,\delta)} w_i = pW + \sum_{i \in M} w_i \ge T = pW+U,
\]
we obtain $\sum_{i \in M} w_i = \sum_{i \in M} b_i \ge U$ and $M$ is a solution to KNAPSACK.
Vice versa, for the same path network any solution $M$ to KNAPSACK can easily be converted into a solution for $p$-Up-MCLP-D by setting $X_p=V^{odd}$ and $\delta_{2k-1} = g_k$, $k \in M$. 

As a result, on path networks $p$-Up-MCLP-D is NP-complete and $p$-Up-MCLP is NP-hard.
\EndProofNoNL
	
The question whether the $p$-Up-MCLP with uniform weights is NP-hard or solvable in polynomial time remains open.

\section{Tree networks}\label{sec:tree}
Theorem~\ref{theo:NPhard:1_facility_stars1_nonuni} shows that the 1-Up-MCLP is NP-hard on weighted star networks. Therefore, the 1-Up-MCLP is NP-hard on weighted tree networks, but analogously to the knapsack problem, it admits a pseudo-polynomial algorithm for the case where the input parameter values are integer.
\medskip
 
\noindent
\textbf{Assumptions}
\begin{description}
 	\item[A1:] The parameters $u_e$ and $\ell_e$ only take non-negative integer values.
 	\item[A2:] The parameters $R$ and $c_e$ only take positive integer values.
\end{description}
 
Let $S=(v_0, \delta)$ be an optimal solution of the 1-Up-MCLP. We call an edge $e \in E$ \emph{fractional} with respect to $S$, if $\delta_e \not\in \mathbb{N}_0:=\mathbb{N} \cup\{0\}$, otherwise it is called  \emph{integer} with respect to $S$. 
If the context is clear, then we usually omit the reference to $S$.
 
\begin{theorem}
\label{theo:1facility:tree:integer_solutions}
If Assumptions A1 and A2 hold, then there exists an optimal solution for the 1-Up-MCLP on trees where all edge upgrades are integer, i.e., $\delta_e \in \mathbb{N}_0$, $e \in E$.
\end{theorem}
 
\ProofNoNL
Let $N=(V,E,\ell)$ be a tree and $S = (v_0, \delta)$ be an optimal solution to the 1-Up-MCLP. If the solution has at least one fractional edge $\be \in E$, then we will show how to construct an equivalent optimal solution with at least one fractional edge less.
In the following we interpret $N$ as a tree rooted at $v_0$. For a node $v_i \in V \setminus \{v_0\}$, $T(v_i)=T_i = (V_i, E_i)$ denotes the sub-tree rooted at $v_i$ consisting of all descendants of $v_i$, including $v_i$ itself, that are covered by $v_0$ with respect to $\delta$. $E_i$ consists of all edges which have both nodes in $V_i$.
 
Let $\be = (v_i, v_j) \in E$ be a fractional edge, i.e., $\delta_\be \not\in \mathbb{N}_0$, such that $i \in T_j$ and there is no other fractional edge on the shortest path from $v_j$ to $v_0$. 
%
We have the following cases to consider:
\begin{description}
\item[Case 1:] All edges in $E_i$ are integer with respect to $S$, i.e., $\delta_e \in \mathbb{N}_0$, $\forall e \in E_i$.
 	
As all edges on the shortest path from $v_j$ to $v_0$ are integer, as well as all edge lengths and $R$, we can round down $\delta_\be$ to the nearest integer without losing coverage of any node in $T_i$, preserving the optimality of $S$ (as $c_\be > 0$).

\item[Case 2:] There is at least one fractional edge in $E_i$.

Let $E_i' \subseteq E_i$ be the set of fractional edges in $E_i$ such that there is no other fractional edge on the path between $e \in E_i'$ and \be. Moreover, let $V_i' \subseteq V_i$ be the set of those nodes of edges $e \in E_i'$ that are further away from $v_0$ than their counterpart on $e$, i.e., $V_i' = \{k \in V_i \mid e=(k,l) \in E_i' \text{ and } d(v_0,v_k) > d(v_0,v_l)\}$.
Finally, set $C = \sum_{e \in E_i'} c_{e}$.
Figure~\ref{fig:1facility:tree:integer_solutions:Case2} depicts the setting with $E_i'=\{ \hat e, \tilde e\}$ and $V_i'=\{v_5,v_8\}$.

\begin{figure}[htb]
\centering
\resizebox{.95\linewidth}{!}{
\begin{tikzpicture}
	\node[scale=.8,auto=left,style=circle,draw] (n1) at (0.5,1.5) {$v_1$};
 	\node[scale=.8,auto=left,style=circle,draw] (n2) at (0.5,-1.5) {$v_2$};
 	\node[scale=.8,auto=left,style=circle,draw] (n0) at (2,0)  {$v_0$};
 	\node[scale=.8,auto=left,style=circle,draw] (nj) at (4.0,0.5)  {$v_j$};
 	\node[scale=.8,auto=left,style=circle,draw] (n3) at (5.5,2)  {$v_3$};
 	\node[scale=.8,auto=left,style=circle,draw] (ni) at (5.5,-1)  {$v_i$};
 	\node[scale=.8,auto=left,style=circle,draw] (n4) at (7.3,0) {$v_4$};
 	\node[scale=.8,auto=left,style=circle,draw] (n7) at (8.5,1.5)  {$v_7$};
 	\node[scale=.8,auto=left,style=circle,draw] (n5) at (8.5,-1.5)  {$v_5$};
 	\node[scale=.8,auto=left,style=circle,draw] (n6) at (10.0,-1) {$v_6$};
 	\node[scale=.8,auto=left,style=circle,draw] (n8) at (10.5,1) {$v_8$};
 			
 	\draw (n0) -- (n1);
 	\draw (n0) -- (n2);
 	\draw (n0) -- (nj);
 	\draw (nj) -- (n3);
 	\draw (nj) -- (ni) node[pos=0.5, above]{$\bar e$};
 	\draw (ni) -- (n4);
 	\draw (n4) -- (n7);
 	\draw (n4) -- (n5) node[pos=0.5, above]{$\hat e$};
 	\draw (n5) -- (n6);
 	\draw (n7) -- (n8) node[pos=0.5, above]{$\tilde e$};
\end{tikzpicture}
}
\caption{Illustration of the notation used in Case 2 of Theorem~\ref{theo:1facility:tree:integer_solutions}.}
\label{fig:1facility:tree:integer_solutions:Case2}
\end{figure}
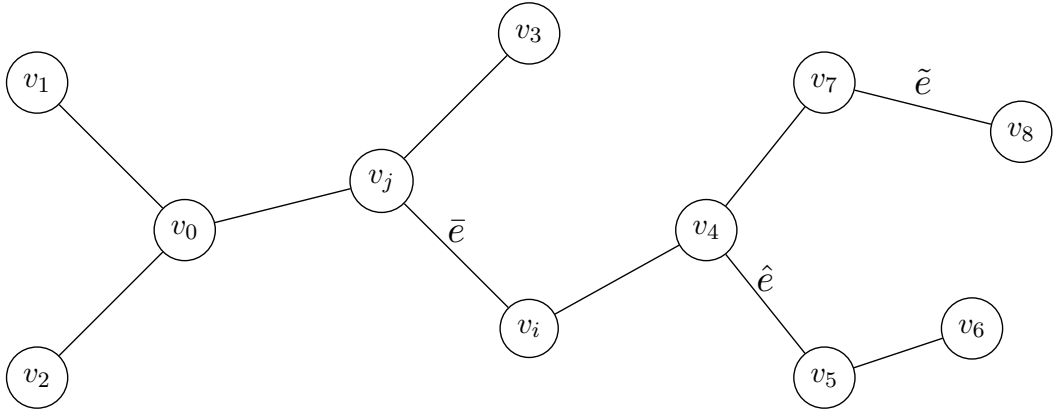

We start with the case that $c_\be \le C$. If we increase $\delta_\be$ by a small amount $\epsilon$ and decrease every $\delta_e$, $e \in E_i'$, by the same amount, then all nodes remain covered: The distance to $v_0$ will decrease for every node in $V_i \setminus \bigcup_{k \in V_i'} V_k$ and will remain unchanged for each $v \in \bigcup_{k \in V_i'} V_k$. For example in Figure~\ref{fig:1facility:tree:integer_solutions:Case2}, the distance from $v_0$ to $v_i$, $v_4$, and $v_7$ will decrease, while it will stay the same for $v_5$, $v_6$, and $v_8$.
Let $\epsilon = \min \{\lceil \delta_\be \rceil - \delta_\be,\, \min_{e \in E_i'} (\delta_e - \lfloor \delta_e \rfloor)\}$. As $c_\be \le C$, increasing $\delta_\be$ by $\epsilon$ and reducing $\delta_e$ by $\epsilon$ for each $e \in E_i'$ will not exceed the budget and will preserve coverage with the same objective value, i.e., it will preserve optimality and result in one less fractional edge.

Next, we consider the case $C < c_\be$. If we decrease $\delta_\be$ by a small amount $\epsilon$, then every node in $V_i \setminus \bigcup_{k \in V_i'} V_k$ will remain covered, even if the distance to $v_0$ increases. And if we increase $\delta_e$ for every $e \in E_i'$ by the same amount, then the distance to $v_0$ will be the same for each $v \in \bigcup_{k \in V_i'} V_k$. Using $\epsilon = \min \{\delta_\be - \lfloor \delta_\be \rfloor,\, \min_{e \in E_i'} (\lceil \delta_e \rceil - \delta_e) \}$, we can again preserve optimality while reducing the number of fractional edges by one.

\end{description}
 
This completes the proof.
\EndProofNoNL

\begin{coro}
 	If Assumptions A1 and A2 hold, then there exists an optimal solution for the 1-Up-MCLP on trees where the budget invested in upgrading an edge is always integer.
\end{coro}
 
\ProofNoNL
The result follows immediately from the integrality of the $\delta$-variables and the upgrading costs $c_e$.
\EndProofNoNL

The above two results can be extended to the $p$-facility case on general networks.

\begin{theorem}
 	If Assumptions A1 and A2 hold, then there exists an optimal solution for the $p$-Up-MCLP on general networks, where all edge lengths are integers, i.e., $\delta_e \in \mathbb{N}_0$, $e \in E$, and the budget invested in upgrading an edge is also integer.
\end{theorem}
\ProofNoNL
Let $S=(X_p,\delta)$ be an optimal solution for $p$-Up-MCLP and let $e=(i,j)$ be fractional after the upgrade.

First, we assume that $i$ and $j$ can both be covered by the same facility $x \in X_p$, i.e., $i, j \in C(\{x\},\delta)$. Without loss of generality, let $d(i,x,\delta) \le d(j,x,\delta)$. We distinguish two different cases:
\begin{itemize}
  \item If no shortest path from $j$ to $x$ passes $i$, then also no shortest path from $x$ to any other node $k \in C(\{x\},\delta)$ will cross $e$. Hence, we can reduce $\delta_e$ by a small amount $\epsilon$ without affecting the distance from $x$ to any  $k \in C(\{x\},\delta)$ covered by $x$. Therefore, we can replace $\delta_e$ by $\lfloor \delta_e \rfloor$ and we have one fractional edge less.
  \item If a shortest path from $j$ to $x$ passes $i$, then we compute the shortest path tree $T_x$ rooted at $x$ for all nodes in $C(\{x\},\delta)$. $T_x$ must necessarily include $e$. Using the same argumentation as in the proof of Theorem~\ref{theo:1facility:tree:integer_solutions}, we can reduce the number of fractional edges by one.
\end{itemize} 

Next, we assume that $i$ and $j$ are covered by distinct facilities $x$ and $x'$, respectively, $x, x' \in X_p$. Without loss of generality, we assume that $x$ and $x'$ are the closest facilities to $i$ and $j$, respectively. As $i \in C(\{x\},\delta)$ but $j \not\in C(\{x\},\delta)$, a shortest path from $i$ to $x$ cannot cross $j$, and vice versa a shortest path from $j$ to $x'$ cannot cross $i$. Thus, edge $e$ is not included in the shortest path trees rooted at $x$ or $x'$ as described above, or, in fact, in the shortest path trees rooted at any facility in $X_p$. Thus, we can reduce $\delta_e$ by a small amount $\epsilon$ without affecting the distance from any node $k \in V$ to its closest facility in $X_p$. Replacing $\delta_e$ by $\lfloor \delta_e \rfloor$, we have one fractional edge less in our solution.

This completes the proof.
%
\EndProofNoNL


Based on the previous results, a pseudo-polynomial algorithm to solve the 1-Up-MCLP on trees is presented. Let $N=(V,E,\ell)$ be a tree 
rooted at node $v_0 \in V$. We follow the construction in \cite{Tam96} to convert $N$ into an equivalent binary tree, i.e., each node has exactly two descendants, as follows. For each non-leaf node $v$ with only one descendant, we introduce a new node $a$ and a new edge connecting $v$ and $a$. $a$ has weight 0, $\ell_{(v,a)}=u_{(v,a)}=0$, and $c_{(v,a)}=1$. For each non-leaf node $v$ with $r \ge 3$ descendants, $b_1,\ldots,b_r$, we introduce $r-2$ new nodes $a_2, \ldots, a_{r-1}$ with weight 0. We remove the edges between $v$ and $b_2, \ldots, b_r$ and replace them with $2r-3$ new edges: $(v,a_2)$, $(a_s,a_{s+1})$, $s=2,\ldots,r-2$, $(a_s,b_s)$, $s=2,\ldots,r-1$, and $(a_{r-1},b_r)$. The first two groups of edges, i.e., $(v,a_2)$ and $(a_s,a_{s+1})$, have length 0, upgrading cost 1, and an upgrading bound of 0. The last two groups of edges, i.e., $(a_s,b_s)$ and $(a_{r-1},b_r)$, have the same parameters as the removed edges $(v,b_s)$, $s=2,\ldots,r$. The resulting tree is binary and has at most $2n-3$ nodes. Moreover, it has the same optimal solution as $N$.

Let Assumptions A1 and A2 hold. In the following, we introduce a leaves-to-root dynamic programming algorithm to determine for a given location $v_0$ of the facility the optimal edge upgrades. For $n \le m$, we use the notation $[n:m]$ to denote the set of integers between $n$ and $m$, including $n$ and $m$.
 
Let $v_0$ be the facility location and also the root of the tree. We denote by $L \subset V$ the set of leaves of the tree. For every $v_i \in V \setminus L$, we denote $v_i^l$ and $v_i^r$ the left and right child, respectively, of $v_i$. The corresponding edges are denoted as $e_i^l$ and $e_i^r$, respectively. For simplicity, let $\ell_i^l = \ell_{e_i^l}$, $\delta_i^l = \delta_{e_i^l}$ and $c_i^l = c_{e_i^l}$; analogously for the right-hand child.
 
At the core of the algorithm is the function
\[ f: V \times \N_0 \times \N_0 \:\to\: \Real,\, (v,d,b) \mapsto f(v,d,b)
\]
which gives the maximal amount of demand that can be covered in the subtree $T(v)$ rooted at $v$ if the distance from $v$ to $v_0$ is $d$ and the budget for upgrading edges in $T(v)$ is $b$.
Obviously, we must have $b \in [0:B]$.
Moreover, if $d > R$, then none of the nodes in $T(v)$ can be covered, including $v$, so the covered demand is zero. 
 
Starting with the leaves, let $v_i \in L$, $b \in [0:B]$, and $d \in [0:d(v_i,v_0)]$. We have
\[ f(v_i,d,b) \:=\: \begin{cases}
 	w_i, & \text{ if } d \le R, \\
 	0, & \text{ otherwise.}
\end{cases}
\]
 
Let now $v_i \in V \setminus L$ be a non-leaf node. For $d \le R$, the optimal coverage in $T_i$ can be calculated as
\[ f(v_i,d,b) \:=\: w_i \:+
\max_{b_i^l + b_i^r + c_i^l\,\delta_i^l + c_i^r\,\delta_i^r = b \atop {\delta_i^l \in [0:u_i^l], \delta_i^r \in [0:u_i^r] \atop b_i^l, b_i^r \in [0:b]}} \:
f(v_i^l,\, d + \ell_i^l - \delta_i^l,\, b_i^l) \:+\: f(v_i^r,\, d + \ell_i^r - \delta_i^r,\, b_i^r) \,.
\]
The maximum considers all possible combinations of splitting the budget $b$ between the left and right subtree, $b_i^l$ and $b_i^r$, and the two edges $e_i^l$ and $e_i^r$, connecting $v_i$ with its children. For $d > R$, $f(v_i,d,b)$ is zero.
 
The optimal coverage is then given by $f(v_0,0,B)$.  We summarise the steps in Algorithm~\ref{algo:tree_v2} and provide its complexity in Theorem~\ref{theo:complexity:1_facility_tree}.
\medskip
 
%
%
%
%
%
%
%

\begin{algorithm2e}[htb]
 	\DontPrintSemicolon
 	\SetAlFnt{\small\sl} 
 	\SetAlCapFnt{\small\sl} \AlCapFnt 
 	
 	\caption{Optimal algorithm for the single facility Up-MCLP on trees with integer parameters.}
 	\label{algo:tree_v2}
 	\BlankLine
 	
 	\KwIn{Tree network $N=(V,E,\ell)$; bounds $u_e$ and upgrade costs $c_e$, for $e\in E$; coverage radius $R > 0$; budget $B>0$.}
 	\BlankLine
 	\KwOut{Optimal solution $(v^*, w^*, \delta^*)$.}
 	\BlankLine \BlankLine
 	
 	Set $w^* = 0$.
 	\BlankLine
 	
 	\ForEach{$v_0 \in V$}{
 	    Determine the binary tree rooted at $v_0$. \;
 	    \ForEach{$v_i \in L, \:b \in [0:B], \:d \in [0:\min\{d(v_i,v_0),R\}] $}{
 	     Define $f(v_i,d,b) \:=\: w_i.$\; 
 	    }
 	     \ForEach{$v_i \in V\setminus L, \:b \in [0:B], \:d \in [0:\min\{d(v_i,v_0),R\}]$}{
 	    	 Calculate  
 	     \[ f(v_i,d,b) \:=\: w_i \:+
 \max_{b_i^l + b_i^r + c_i^l\,\delta_i^l + c_i^r\,\delta_i^r = b \atop{\delta_i^l \in [0:u_i^l], \delta_i^r \in [0:u_i^r] \atop b_i^l, b_i^r \in [0:b]}} \:
 f(v_i^l,\, d + \ell_i^l - \delta_i^l,\, b_i^l) \:+\: f(v_i^r,\, d + \ell_i^r - \delta_i^r,\, b_i^r) \,.
 \]}
 	     
 			\If{$w=f(v_0,0,B) > w^*$}{
 				Set $w^* = w,$ $\delta^* = \delta,$ and $v^* = v_0$.
 			}
 		
 	}
 	
 	\BlankLine
 	
 	\Return\ $(v^*,w^*, \delta^*)$.\;
\end{algorithm2e}

\begin{theorem} 	\label{theo:complexity:1_facility_tree}
     The 1-Up-MCLP on tree networks with integer parameters can be solved in $O(|V|^2RB^3)$ time using Algorithm~\ref{algo:tree_v2}.
\end{theorem}
  
\ProofNoNL
Let $v_0$ be fixed. The core of the algorithm consists of computing the function $f$. Step~5 is computed in constant time for each $v_i \in L, \:b \in [0:B], \:d \in [0:\min\{d(v_i,v_0),R\}]$. 

Let $v_i \in V \setminus L$ be a non-leaf node, $d \in [0:R]$, and $b \in [0:B]$. There are $b+1$ ways to split the budget between the left and right child: $(\bar b_i^l, \bar b_i^r) \in \{(b,0), (b-1,1), \ldots, (0,b)\}$. The budget $\bar b_i^l$ for the left child can be spent on upgrading the edge $e_i^l$ and/or upgrading edges in the subtree $T(v_i^l)$. The number of combinations depends on $c_i^l$ and $u_i^l$. We have that $\delta_i^l \le \min \{u_i^l, \lfloor \bar b_i^l / c_i^l \rfloor \}$, which is bounded by $b$. Considering all possible values for $\delta_i^l$ for each value of $b_i^l$, we have in total at most
\[ (b+1) + b + (b-1) \ldots + 1 = O(b^2) = O(B^2)
\]
combinations to consider for the left child. Analogously for the right child. Note that whenever we call $f(\cdot)$, it takes constant time to check whether or not $d + \ell_i^l - \delta_i^l \le R$ ($d + \ell_i^r - \delta_i^r \le R$). Therefore, the complexity of computing Step~8 is $O(B^2).$ Step~3 takes $O(|V|)$ time \citep{Tam96} and Step~11 takes constant time. 
 
Therefore, going over all possible combinations of $d$ and $b$ for Step~8, the total effort to work through all combinations for a given node $v_i$ is $O(RB^3)$. Doing this for every node, the overall complexity for finding the optimal upgrade for a given facility location $v_0$ is $O(|V|RB^3)$.

Finally, testing all nodes $v_0 \in V$ as facility locations, we can solve the 1-facility upgrading problem on trees with integer parameters in $O(|V|^2RB^3)$ time. Note that we do not have to construct the binary tree every time from scratch. Once we have obtained the equivalent binary tree for a node $v_0 \in V$, we can incrementally update it for a new root $v_0' \in V$. For that we have to add a one new node $a_0$ and connect it to $v_0$, and another new node $a_0'$ and connect it to $v_0'$ and two of its descendants, replacing the edges between $v_0'$ and those descendants (provided that none of the descendants is a node that was added in the initial transformation, in which case we can simply delete this node and its incident edge). The attributes of the new nodes and edges are set in the same way as above. This can be done in constant time and the total number of nodes and edges of the binary tree increases by at most two and, respectively, one.
\EndProofNoNL
 

In the following, we provide a small example of solving an instance using Algorithm~\ref{algo:tree_v2}.
\begin{example}
    Consider the network depicted in Figure~\ref{fig:instance}.  For each node, its name is printed inside the node and its weight is next to it. Let
$R=2$, $B=3$, $\ell_j=2,$ and $c_j=u_j=1$, for $j\in E$.

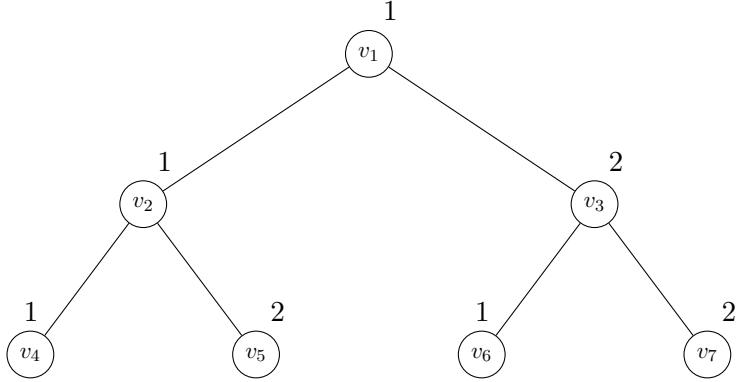
\begin{figure}[htb]
 	\centering
 	\begin{tikzpicture}
 		
 	\node[scale=.8,auto=left,style=circle,draw,label=  80:$1$] (n3) at (4,0)  {$v_1$} ;
 		\node[scale=.8,auto=left,style=circle,draw,label=  80:$1$] (n4) at (1,-2)  {$v_2$};
 		\node[scale=.8,auto=left,style=circle,draw,label=  80:$2$] (n5) at (7,-2)  {$v_3$};
 		\node[scale=.8,auto=left,style=circle,draw,label=  88:$1$] (n6) at (-0.5,-4) {$v_4$};
 		\node[scale=.8,auto=left,style=circle,draw,label=  80:$2$] (n7) at (2.5,-4) {$v_5$};

   \node[scale=.8,auto=left,style=circle,draw,label=  88:$1$] (n8) at (5.5,-4) {$v_6$};
 		\node[scale=.8,auto=left,style=circle,draw,label=  80:$2$] (n9) at (8.5,-4) {$v_7$};
   
 		\draw (n3) -- (n4) ;
 		\draw (n3) -- (n5) ;
 		\draw (n4) -- (n6) ;
 	      \draw (n4) -- (n7) ;
            \draw (n5) -- (n8) ;
            \draw (n5) -- (n9) ;
 	\end{tikzpicture}
 	\caption{Binary tree with node weights in brackets.}
  \label{fig:instance}
\end{figure}
We will describe the procedure for obtaining the optimal upgrading strategy assuming the facility's location is node $v_1$ (i.e., $v_1$ takes the role of $v_0$ using the notation of the algorithm). Observe that the algorithm requires that we repeat this process for each node of the tree in order to obtain the optimal location of the facility. Then, we will determine which facility location covers more demand after applying the optimal upgrading strategy.  

We have $L=\{v_4,v_5,v_6,v_7\}$ as the set of leaves of the tree. For $v_i \in L,\:d \in [0:2],  \:b \in [0:3]$, we have $f(v_i,d,b) \:=\: w_i.$ For other values of $d$ and $b$ this function has value zero. 

In the following, Steps 7 and 8 of Algorithm \ref{algo:tree_v2} shall be carried out for $v_2$ and $v_3$. To exemplify, we describe the process for $v_2$ (it is done analogously for $v_3$). We need to compute $f(v_2,d,b)$  for  $b\in[0:3]$ and $d\in[0:2],$ i.e., 12 different combinations. Note that some combinations of parameters may not be possible to occur for a node, e.g., for $v_2$ we cannot have $d=0$. To illustrate the procedure, we solve one of the 12 combinations, namely $d=1$ and $b=2$:
\[ f(v_2,1,2) \:=\: 1 \:+
\max_{b_2^l + b_2^r + \delta_2^l + \delta_2^r = 2  \atop{\delta_2^l \in [0:1],\, \delta_2^r \in [0:1] \atop b_2^l, b_2^r \in [0:2]}} \:
f(v_4,\, 3 - \delta_2^l,\, b_2^l) \:+\: f(v_5,\, 3- \delta_2^r,\, b_2^r) \,.
\]
Therefore, the optimal solution of the previous problem is $\delta_2^l=\delta_2^r=1, b_2^l=b_2^r=0,$ i.e., 
\[f(v_2,1,2) \:=\:w_2+f(v_4,2,0)+f(v_5,2,0)=w_2+w_4+w_5=4.\]
After performing the same procedure for $v_3$, we repeat the process for $v_1$ and obtain that an optimal upgrading strategy for this location of the facility is $\delta_1^l=1, \delta_2^l=1,$ $\delta_2^r=1$ and zero on all other edges (see Figure~\ref{fig:instance2}), then,
\begin{eqnarray*}
f(v_1,0,3)&=&w_1+f(v_2,1,2)+f(v_3,2,0)=w_1+f(v_4,2,0)+f(v_5,2,0)+w_3\\&=&w_1+w_2+w_3+w_4+w_5=7.
\end{eqnarray*}
 \begin{figure}[htb]
 	\centering
 	\begin{tikzpicture}
 		
 	\node[scale=.8,auto=left,style=circle,draw, label=80:$1$] (n3) at (4,0)  {$v_1$} ;
 		\node[scale=.8,auto=left,style=circle,draw,label=80:$1$] (n4) at (1,-1)  {$v_2$};
 		\node[scale=.8,auto=left,style=circle,draw,label=80:$2$] (n5) at (7,-2)  {$v_3$};
 		\node[scale=.8,auto=left,style=circle,draw,label=145:$1$] (n6) at (-0.5,-2) {$v_4$};
 		\node[scale=.8,auto=left,style=circle,draw,label=35:$2$] (n7) at (2.5,-2) {$v_5$};

   \node[scale=.8,auto=left,style=circle,draw,label=88:$1$] (n8) at (5.5,-4) {$v_6$};
 		\node[scale=.8,auto=left,style=circle,draw,label=80:$2$] (n9) at (8.5,-4) {$v_7$};
   
 		\draw (n3) -- node[above=3mm]{$\delta_1^l=1$} (n4);
 		\draw (n3) -- (n5) ;
 		\draw (n4) -- node[above left]{$\delta_2^l=1$} (n6) ;
 	      \draw (n4) -- node[above right]{$\delta_2^r=1$}(n7) ;
            \draw (n5) -- (n8) ;
            \draw (n5) -- (n9) ;
 	\end{tikzpicture}
 	\caption{Optimal upgrading strategy for locating the facility at $v_1$}
  \label{fig:instance2}
\end{figure}
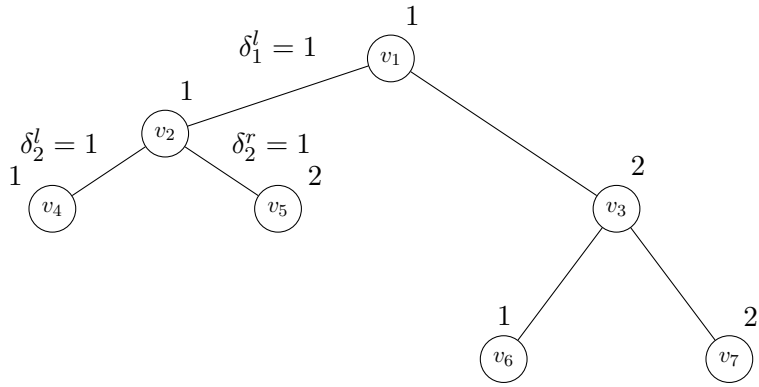
Now, the described procedure has to be repeated assuming that the facility is located at each of the other nodes in the tree (i.e., $v_2, v_3, v_4, v_5, v_6,$ and $v_7$). Note that to do so, an equivalent binary tree rooted at each of the nodes has to be built. Finally, it can be observed that the solution described above is an optimal solution to the problem. 

\end{example}

\bigskip
 
 In this section, we have developed a pseudo-polynomial algorithm for the 1-facility upgrading problem on trees with integer parameters. The remaining open question is whether the $p$-Up-MCLP with uniform weights is NP-hard or polynomial. 
 
\section{Concluding remarks} \label{sec:conclusion}
In this paper, we propose an algorithmic approach to deal with the upgrading version of the maximal covering location problem assuming that the edge length of the edges can be modified. 
 
We prove that the  problem on star networks is polynomial time solvable for uniform weights ($O(n\log n)$) and  NP-hard for non-uniform weights.  On path, we show that the single facility problem is solvable in polynomial time ($O(n^3)$), while the $p$-facility problem is NP-hard even with uniform costs and upper bounds, as well as, integer parameter values. 
Furthermore, a pseudo-polynomial algorithm is developed for the single facility problem on trees with integer parameters ($O(|V|RB^3)$). A summary of the obtained result can be found in Table~\ref{tab:summary}.
 
One of the remaining open question is whether the upgrading $p$-facility maximal covering problem with uniform weights is NP-hard or polynomial on path and tree networks.

\section*{Compliance with Ethical Standards}

\subsection*{Funding}

All authors were partially supported by research project PID2020-114594GB-C22 (Agencia Estatal de Investigaci\'on, Spain and the European Regional Development's funds (ERDF)) and research project RED2018-102363-T (Spanish Ministry of Science and Innovation). 

\textbf{Marta Baldomero-Naranjo} was partially supported by MCIN/AEI/10.13039/501100011033 and the European Union ``NextGenerationEU"/PRTR through project TED2021-130875B-I00 
and the BritishSpanish Society Scholarship Programme 2018 (Telef\'onica and the BritishSpanish Society). 

\textbf{Jörg Kalcsics} was partially supported by Grant number EST2019-047 for research stays in Universidad de Cádiz (Plan Propio de Investigación y Transferencia de la UCA). 

\textbf{Antonio M. Rodr\'iguez-Ch\'ia} was partially supported by MCIN/AEI/10.13039/501100011033 and the European Union ``NextGenerationEU"/PRTR through project TED2021-130875B-I00.    

The authors would like to thank the anonymous reviewers for their comments and suggestions and declare that they have no conflict of interest.

\subsection*{Ethical approval}
This article does not contain any studies with human participants or animals performed by any of the authors.

\bibliographystyle{abbrvnat}
\bibliography{./references}

\end{document}